# Stacking-Engineered Thermal Transport and Phonon Filtering in Rhenium Disulfide

Yongjian Zhou[1] ‡, Haoran Cui[2] ‡, Zefang Ye[1] ‡, Jung-Fu Lin[3], Yan Wang[2]* and Yaguo Wang[1]*

1. Walker Department of Mechanical Engineering, The University of Texas at Austin, Austin, TX, 78712, USA

2. Department of Mechanical Engineering, University of Nevada, Reno, Reno, NV 89557, USA

3. Department of Geological Sciences, Jackson School of Geosciences, The University of Texas at Austin, 2305 Speedway Stop C1160, Austin, Texas 78712, USA

E-mails: yaguo.wang@austin.utexas.edu, yanwang@unr.edu.

‡These authors contributed equally.

ABSTRACT: Cross-plane heat transport is a critical bottleneck for van der Waals (vdW) electronics, yet its microscopic governing principles remain elusive. We demonstrate that stacking order is an effective control knob for cross-plane phonon transport in multilayer

Rhenium Disulfide (ReS$_2$). Thickness-dependent thermal conductivity measurements reveal remarkably long cross-plane phonon mean free paths (MFPs) ($\gtrsim$ 200–300 nm) and provide a direct experimental observation of the transition from quasi-ballistic transport to a thickness-independent ballistic limit. AA stacking exhibits nearly double the cross-plane thermal conductivity of AB stacking, driven by longer acoustic phonon lifetimes from a more "coherent" interlayer registry. Integrated deep neural-network molecular dynamics reveals that phonon filtering in ReS$_2$ is fundamentally frequency-selective: weak vdW coupling acts as a low-pass filter, whereas stronger coupling broadens the transmission passband. These results establish ReS$_2$ as a model system where stacking order and interlayer coupling can be engineered to tune heat conduction across diffusive, quasi-ballistic, and ballistic regimes, offering a new framework for thermal management in 2D electronics.

ReS$_2$ has emerged as a versatile two-dimensional (2D) material that is widely investigated for photodetectors, ultrafast saturable absorbers, and other optoelectronic applications owing to its high photoresponsivity, broadband absorption, and strong nonlinear optical response. [1-3] ReS$_2$ crystallizes in a distorted 1T triclinic structure in which rhenium atoms form zigzag chains. This low-symmetry lattice dramatically enhances in-plane anisotropy and, when combined with weak interlayer vdW coupling, endows ReS$_2$ with physical properties that are highly sensitive to structural registry. Multilayer ReS$_2$ can adopt two well-defined stacking orders: a "coherent" AA registry with perfectly aligned layers, and an AB configuration defined by a half-unit-cell lateral shift, which correspond to distinct local energy minima within the crystal lattice. [4] These stacking configurations exert strong influence on the material's vibrational, optical, and electronic responses. As such, stacking order constitutes an additional and independent structural

degree of freedom for tuning the properties of ReS$_2$, beyond layer number and in-plane orientation. [5-7]

Similar to other vdW materials, ReS$_2$ exhibits strongly anisotropic thermal transport, with in-plane thermal conductivity (~70 W m$^{-1}$ K$^{-1}$ along Re-chain and ~50 W m$^{-1}$ K$^{-1}$ perpendicular to Re-chain) substantially exceeding its cross-plane counterpart (~ 0.55 W m$^{-1}$ K$^{-1}$ in nanofilms), as reported by prior time-domain thermoreflectance (TDTR) measurements.[8] In vdW-based electronic and optoelectronic devices, heat generated in the active layers dissipates predominantly along the cross-plane direction through stacked layers and interfaces, making cross-plane thermal conductivity ($\kappa_c$) a critical bottleneck for device performance and reliability.[9] Understanding how interlayer registry controls cross-plane phonon transport is therefore essential for engineering next-generation vdW electronics and optoelectronics. Historically, cross-plane thermal transport in layered vdW materials was assumed to be dominated by phonons with very short MFPs, typically on the order of only a few nanometers. This picture has been fundamentally revised by recent experimental and theoretical studies on benchmark systems such as graphite [10-12] and MoS$_2$ [13], which demonstrate that room-temperature cross-plane phonon MFPs can extend to hundreds of nanometers. Such unexpectedly long-range transport has been attributed to a phonon filtering effect, in which weak interlayer vdW interactions suppress short-MFP phonons and confine heat flow to long-MFP modes.

ReS$_2$ presents a unique and largely unexplored opportunity to examine cross-plane phonon transport in exceptional depth. Because its AA and AB stacking sequences remain remarkably stable across varying thicknesses, ReS$_2$ serves as an ideal platform for investigating and ultimately engineering cross-plane heat transport in ultrathin vdW devices. While earlier TDTR measurements reported no clear thickness dependence for ReS$_2$ films between 60 nm and 450 nm, these studies did not distinguish between stacking orders and therefore effectively averaged over fundamentally different interlayer configurations. [8] Consequently, the role of interlayer registry in governing phonon filtering, phonon MFPs, and the crossover between distinct transport regimes remains a critical and practically important open question.

In this work, we perform systematic picosecond transient thermoreflectance (ps-TTR) measurements of high-quality ReS$_2$ samples with well-defined AA or AB stacking orders that persists over micrometer-scale thicknesses. These measurements are thus capable of capturing an order-or-magnitude variation in $\kappa_c$ with increasing sample thickness, as well as pronounced differences between AA- and AB-stacked ReS$_2$. By integrating the ps-TTR measurements with molecular dynamics simulations based on a deep neural network (DNN) potential trained from ab initio, we further compare the spectral phonon properties of AA- and AB-stacked ReS$_2$ for both ambient pressure and high-pressure conditions. Through this combined experimental-computational approach, we elucidate the unique phonon transport behavior of ReS$_2$ and demonstrate it as a model system in which stacking order, interlayer coupling, and phonon filtering can be deliberately manipulated to tune cross-plane heat conduction from diffusive to quasi-ballistic and fully ballistic regimes.

Figs. 1a and 1b show the atomic configurations of the AA and AB stacking orders in $ReS_2$. The stacking order of $ReS_2$ samples can be easily identified by measuring the difference between two Raman peaks: $\Delta$ = mode III - mode I. For AA stacking, $\Delta \sim 13$ cm$^{-1}$, and for AB stacking, $\Delta \sim 20$ cm$^{-1}$. We employed a re-exfoliation approach to confirm that a single, well-defined stacking order can persist over thicknesses of several micrometers. (Fig. S1 & S2) Cross-plane thermal conductivities were measured using a ps-TTR technique [14, 15]. The thickness-dependent $\kappa_c$ for both AA- and AB-stackings are presented in Fig. 1c. In the ultrathin regime (below 100 nm), both stacking configurations exhibit similar values, which is consistent with prior measurements [8] and suggests that thermal transport in the thinnest flakes is primarily constrained by boundary scattering. However, as thickness increases, the thermal behaviors of the two registries diverge significantly. While both configurations show a pronounced increase in $\kappa_c$ with thickness, AA-stacked samples exhibit a more rapid rise and reach a clear saturation plateau of approximately 4.0 W m$^{-1}$ K$^{-1}$. In contrast, AB stacking exhibits a slower recovery of $\kappa_c$ and fails to reach saturation even as thicknesses approach 1 μm, trending toward an extrapolated bulk limit of approximately 2.0 W m$^{-1}$ K$^{-1}$. This distinct bifurcation demonstrates that stacking order serves as a primary determinant of intrinsic cross-plane transport, while the delayed saturation in both registries provides direct experimental evidence of exceptionally long cross-plane phonon MFPs that far exceed the predictions of conventional kinetic theory.

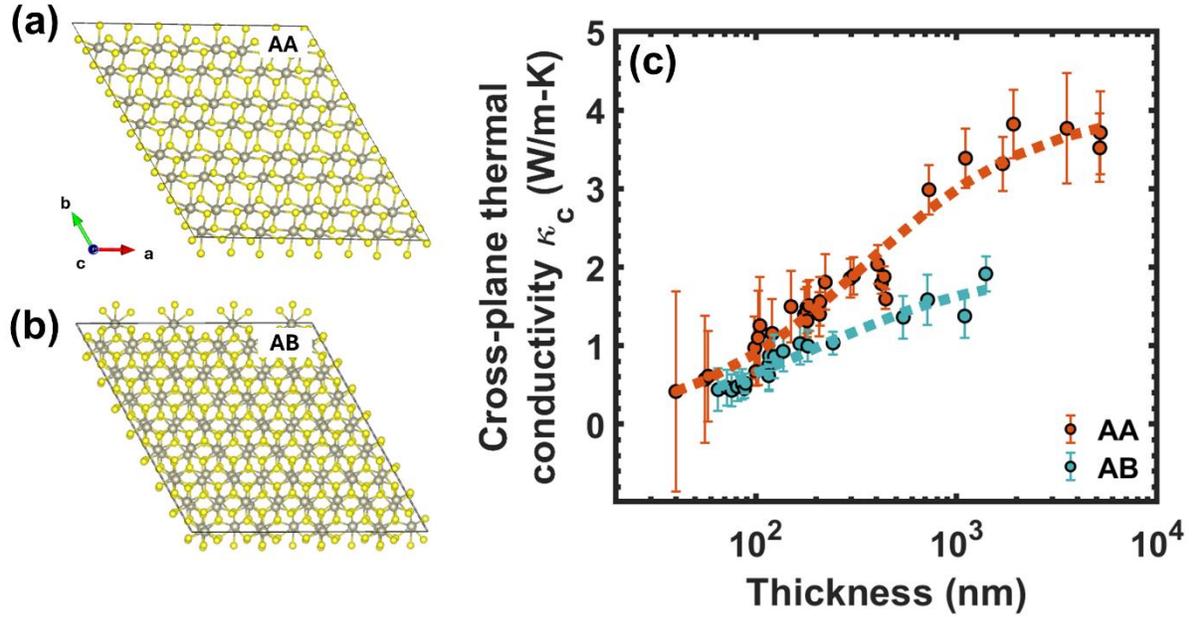

**Figure 1**. (a,b) Atomic structures of AA and AB stacking orders, where grey dots represent Re atoms and yellow dots represent S atoms. (c) Measured cross-plane thermal conductivity as a function of thickness for AA- and AB-stacked ReS$_2$. Dashed lines represent fittings with gray model.

To quantify the characteristic MFPs, we fit the experimental $\kappa_c$ using the gray model [16, 17],

$$\kappa_{film} = \kappa_{bulk} \frac{d}{d + \Lambda_{bulk}}$$

The extracted bulk $\kappa_c$ values agree with the experimental saturation trends: ~4.0 W m$^{-1}$ K$^{-1}$ for AA stacking and ~2.0 W m$^{-1}$ K$^{-1}$ for AB stacking. The corresponding bulk phonon MFPs are remarkably long, ~346 nm for AA and ~203 nm for AB. If we apply the kinetic expression: $\kappa_c \sim 1/3\, C\boldsymbol{v}_c\Lambda_c$, where $C$ is heat capacity and $\boldsymbol{v}_c$ is cross-plane sound velocity, yields $\Lambda_c \approx 2.2$ nm for AA stacking, corresponding to only ~8 layers. The experimental MFPs surpass kinetic theory predictions by nearly two orders of magnitude. This striking discrepancy arises because conventional kinetic theory fails to distinguish between the short-MFP modes that dominate heat capacity and the long-wavelength phonons that drive thermal transport. By using the total heat capacity, the kinetic model systematically underestimates the effective MFPs of the primary heat-carrying phonons. Resolving these mode-dependent characteristics requires a detailed spectral analysis.

Fig. 1 highlights two critical findings: first, $\kappa_c$ is highly sensitive to stacking order, with AA stacking conducting heat nearly twice as efficiently as the AB configuration. Second, both registries exhibit exceptionally long phonon MFPs, confirming that cross-plane transport in $ReS_2$ is dominated by long-range modes rather than localized vibrations. These observations raise fundamental questions regarding why AA stacking supports much higher conductivity and what physical mechanism drives such long MFPs, which we address through mode-resolved spectral analysis in the following sections.

To address the first question: why AA-stacking $ReS_2$ conducts heat more efficiently than its AB-stacking, we constructed a DNN interatomic potential, using the DeepMD approach [18], from extensive ab initio data for both stacking orders and performed mode-resolved phonon analysis. The accuracy of the DNN potential was validated through non-equilibrium molecular dynamics (NEMD) simulations [19] performed at 300 K using LAMMPS [20]. As demonstrated in Fig. S4, the NEMD results reproduce the experimentally measured $\kappa_c$ for both stacking orders up to ~300 nm, confirming that the DNN potential reliably captures the size-dependent thermal transport in both types of $ReS_2$.

Bulk-limit thermal conductivities were subsequently obtained using equilibrium molecular dynamics (EMD) [21] and the Green–Kubo formalism [22, 23]. The predicted bulk $\kappa_c$ values are approximately 4.5 W ± 0.12 m$^{-1}$ K$^{-1}$ for AA stacking and 3.2 ± 0.11 W m$^{-1}$ K$^{-1}$ for AB stacking, slightly higher than the experimental values, which is expected because defects and imperfections are present in experimental samples but are absent in simulations. Importantly, both experiments and simulations clearly show that AA-stacking structures possess significantly higher $\kappa_c$ than their AB counterparts, underscoring the crucial role of stacking order in determining thermal transport in ReS$_2$.

To identify the microscopic origin of this stacking-dependent behavior, we performed spectral energy density (SED) analysis [24], which yields phonon dispersions, mode-resolved phonon group velocities, lifetimes and MFPs. Fig. 2a displays the phonon dispersion relations obtained from the SED calculations. The LA branch in AA-stacking ReS$_2$ exhibits a slightly steeper slope compared to AB-stacking ReS$_2$, while the two TA branches are nearly degenerate in AA stacking but clearly split in the AB case, consistent with our density functional theory (DFT) calculations (Fig. S7). The corresponding group velocities (Fig. 2b) show that LA velocities in AA are higher than in AB, whereas AB exhibits higher TA velocities. However, the difference in group velocities is not sufficiently large to account for AA's substantially higher $\kappa_c$.

Figs. 2c and 2d show that both LA and TA phonons in AA-stacking ReS$_2$ possess systematically and notably longer lifetimes than those in AB-stacking configuration, which is the primary origin of the higher $\kappa_c$ in AA stacking. The longer phonon lifetimes of AA-stacking ReS$_2$ originates from the near-degeneracy of its TA modes, which limits the available three-phonon scattering channels due to more stringent energy-conservation conditions. In contrast, the TA-mode splitting in AB-stacking ReS$_2$ opens additional allowed scattering pathways, thereby increasing anharmonic phonon scattering and reducing lifetimes. Due to its relatively longer phonon lifetimes, AA-stacking ReS$_2$ has notably higher contribution to $\kappa_c$ by longer-MFP modes than its AB-stacking counterpart, as revealed by the mode-wise $\kappa_c$ and cumulative $\kappa_c$ data in Fig. 2e. Fig. 2f further illustrates this behavior by comparing the three-phonon emission phase space, where one higher-frequency TA mode splits into two lower-frequency modes. The AB-stacking structure exhibits a markedly larger scattering phase space than its AA-stacking counterpart. This interpretation is consistent with the structural configurations in Fig. 1: AA stacking features perfect vertical alignment between adjacent layers, a more "coherent" interlayer registry, while AB stacking introduces a half-unit-cell lateral shift. This structural mismatch in AB stacking naturally increases phonon scattering across layers and lowers the $\kappa_c$.

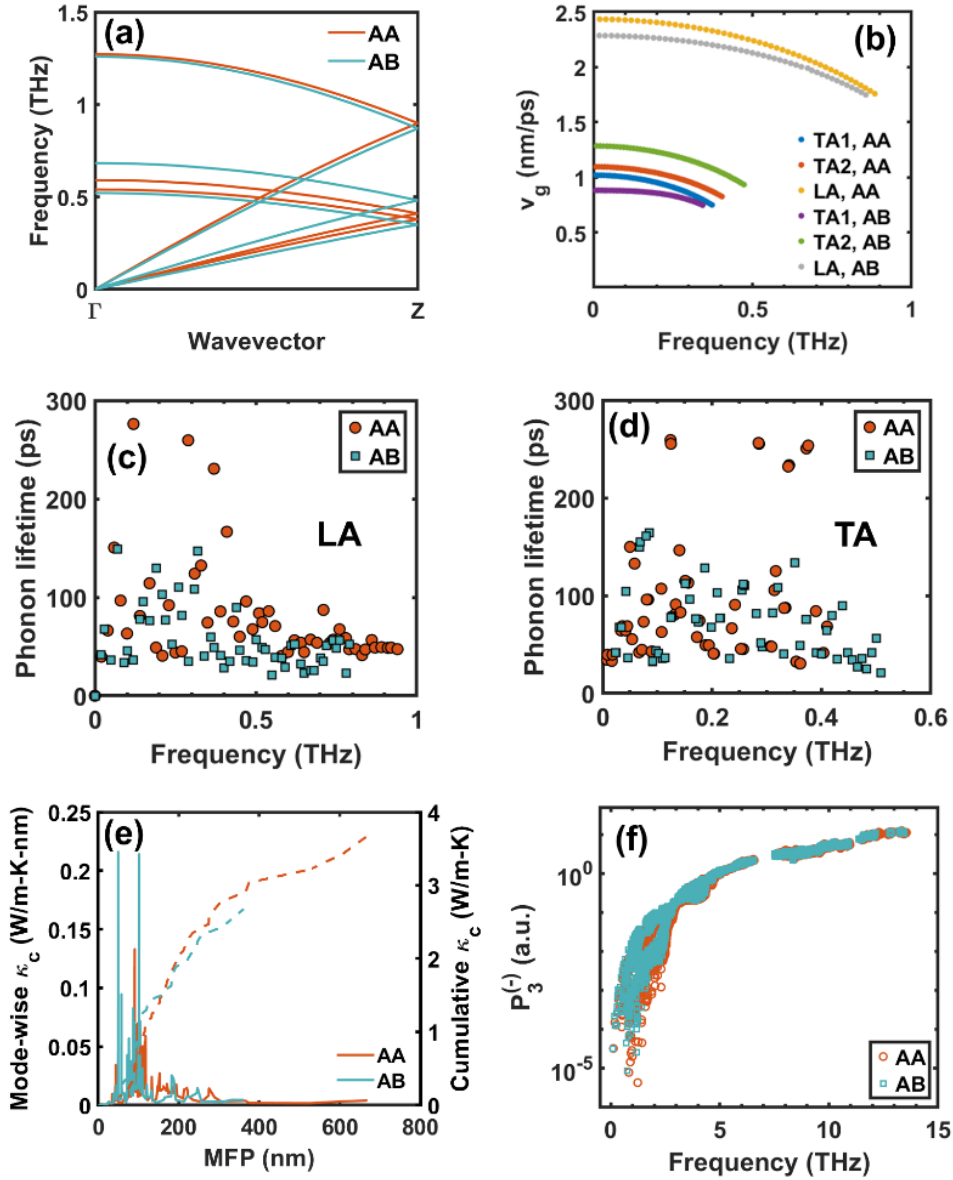

**Figure 2**. Mode-resolved phonon properties of AA- and AB-stacked ReS$_2$ from SED calculations. (a) Phonon dispersion relations for AA and AB stacking. (b) Mode-resolved phonon group velocities. (c,d) Phonon lifetimes for AA and AB stacking extracted from Lorentzian fitting of SED spectra. (e) Left axis: mode-wise cross-plane thermal conductivity. Right axis: cumulative cross-plane thermal conductivity. (f) The emission process of three-phonon scattering phase space for ReS$_2$ with AA and AB stacking orders.

We now address the second question: what governs the long phonon MFP. Similar thickness-dependent measurements on graphite and MoS$_2$ consistently indicate that long-MFP phonons dominate cross-plane heat conduction. In graphite, the c-axis phonon MFP has been estimated to be ~200 nm, [10, 11] while in MoS$_2$, phonons with MFPs exceeding ~200 nm contribute more than 50% of the total $\kappa_c$[12]. These unexpectedly long MFPs have been attributed to the so-called phonon filtering effect, in which the weak interlayer vdW forces suppress short-MFP phonons and preferentially allow long-MFP modes to transmit across layers. If this conventional understanding of phonon filtering effect holds, one would expect the filtering effect to depend strongly on interlayer bonding strength: weaker vdW force → stronger filtering → a larger contribution from longer-MFP phonons (i.e., a spectral shift of mode-wise $\kappa_c$ toward longer MFP), and conversely, stronger interlayer coupling should weaken filtering and contribution from short-MFP phonons should increase (i.e. a spectral shift of mode-wise $\kappa_c$ toward shorter MFP).

Hydrostatic pressure is a convenient tool to tune interlayer force in vdW materials and test this conventional picture of phonon filtering effect. Hydrostatic pressure was applied using a Nosé-Hoover barostat [25, 26] in all three directions. We calculated the cumulative $\kappa_c$ as a function of MFP for both AA and AB stackings at 0 GPa and 20 GPa. As shown in Figs. 3a and 3c, increasing pressure (i.e., strengthening interlayer coupling) shifts the cumulative $\kappa_c$ curves toward longer MFPs for both stackings. For AA stacking, phonons with $\Lambda < 180$ nm contribute 50% of $\kappa_c$ at 0 GPa, but the 50% cutoff increases to ~300 nm at 20 GPa. For AB stacking, the cutoff shifts from ~120 nm at 0 GPa to above 200 nm at 20 GPa. The corresponding mode-wise $\kappa_c$ spectra (Fig. 3b and 3d) further confirm that stronger inter-layer force at higher pressure extends the phonon contribution into the long-MFP regime. This trend arises because applying pressure increases the group velocity $v_c(\omega)$, while phonon lifetimes $\tau(\omega)$ change only weakly (shown in Fig. S8); as a result, the MFP $\Lambda_c(\omega)=v_c(\omega)\tau(\omega)$ becomes longer even though the spectral filter becomes less selective. These observations seem to contradict the conventional understanding of phonon filtering and indicate that the relationship between phonon filtering and MFP requires careful re-examination.

Because phonon filtering is fundamentally a spectral (frequency-domain) phenomenon, we next analyzed cumulative $\kappa$ as a function of phonon frequency (Fig. 4). At 0 GPa, both AA and AB stacking show that ~50% of the $\kappa_c$ arises from phonons around 0.3 THz (Fig. 4a), and virtually all contributing modes lie below ~0.5 THz (Fig. 4b). When pressure increases to 20 GPa, however, the transmission bandwidth broadens dramatically, extending to ~1.5 THz (Fig. 4d and 4f), and the cumulative $\kappa_c$ curves gain substantial weight at higher frequencies (Fig. 4c and 4e). This indicates that weak vdW forces at low pressure act as a strong low-pass *spectral* filter, effectively blocking high-frequency modes and transmitting only long-wavelength phonons.

We quantify the filtering strength using a cutoff frequency $\omega_{cut}$=0.5 THz, defined as $F(\omega_{cut}) = \kappa(\omega<\omega_{cut})/\kappa_{max}$. For AA stacking, F increases from 37.22% at 20 GPa to 78.4% at 0 GPa; for AB stacking, it increases from 35.97% to 80.0% over the same pressure range. These values demonstrate that the filtering strength is controlled predominantly by the vdW interlayer force, rather than the stacking order.

We define the phonon filtering effect as the preferential transmission of low-frequency, long-wavelength modes across the vdW gap, which suppresses high-frequency contributions to cross-plane heat flow. Because this mechanism originates from the wave nature of phonons [27], it is most naturally described in the frequency domain and is closely analogous to optical low-pass filtering. [28] This definition resolves the ambiguity in the literature [29, 30], where the thermal consequences of filtering are often expressed in the MFP domain, since $\Lambda$ determines how far each transmitted mode carries heat. Our results show that these two perspectives must be considered together: phonon filtering is fundamentally a spectral effect, and weaker interlayer coupling narrows the phonon bandwidth (stronger filtering at 0 GPa, Fig. 4), whereas the MFP spectrum reflects the transport outcome of the modes that pass through that filter, resulting in shorter $\Lambda$ at 0 GPa due to lower phonon group velocities (Fig. 3).

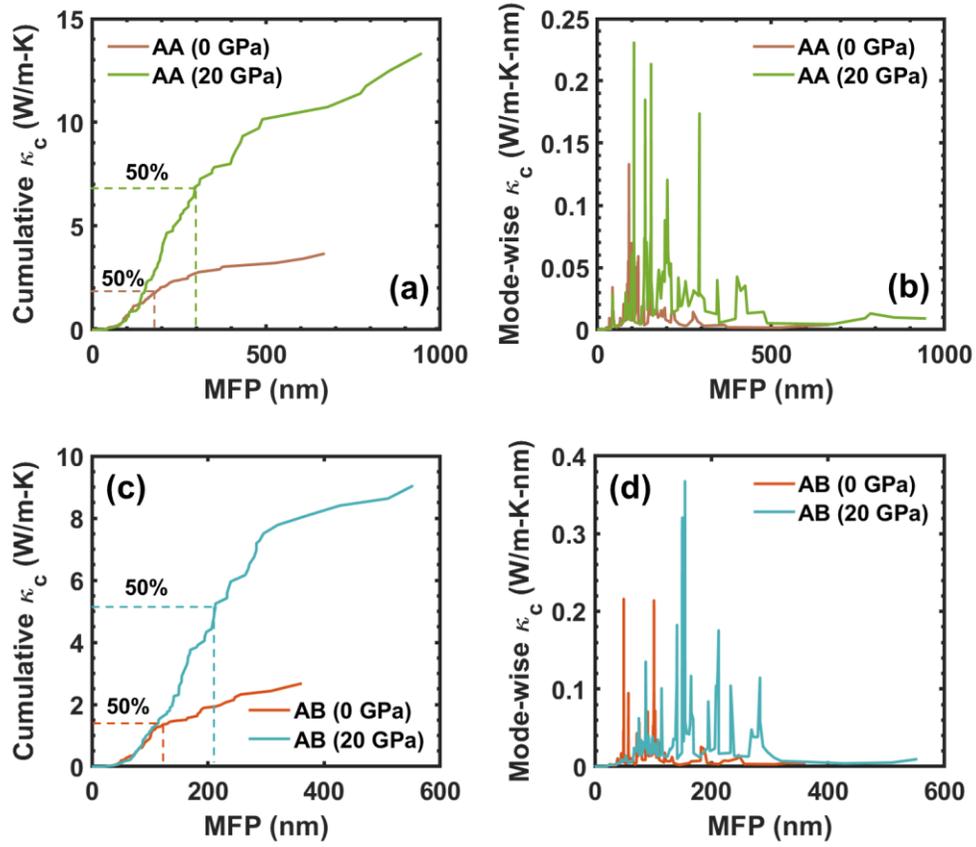

**Figure 3**. *Cumulative cross-plane thermal conductivity as a function of phonon MFP for AA- and AB-stacked ReS$_2$ under different pressures*. (a,c) Cumulative cross-plane thermal conductivity versus MFP at 0 GPa and 20 GPa for AA and AB stacking. (b,d) Corresponding mode-wise cross-plane thermal conductivity spectra displaying the distribution of phonon contributions across MFPs.

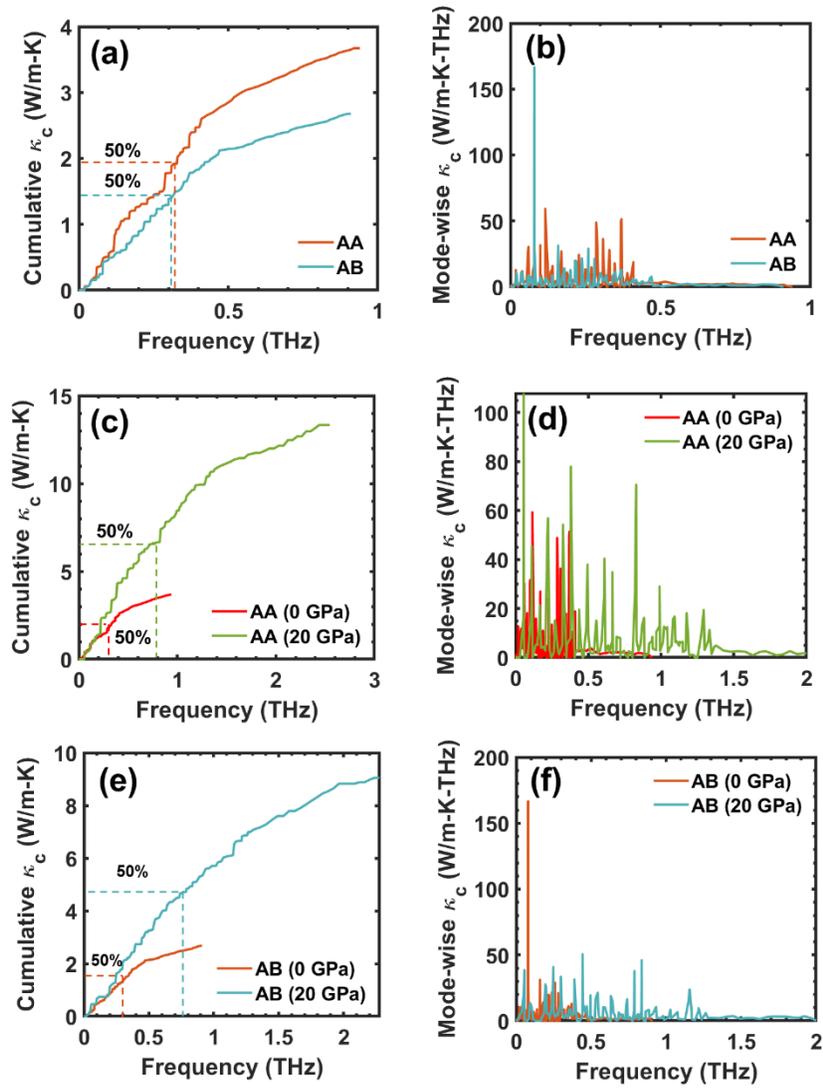

**Figure 4.** *Cumulative cross-plane thermal conductivity as a function of phonon frequency for AA- and AB-stacked ReS$_2$ under different pressures.* (a,c,e) Cumulative cross-plane thermal conductivity versus frequency at 0 GPa and 20 GPa for AA and AB stacking. (b,d,f) Corresponding mode-wise cross-plane thermal conductivity spectra displaying the distribution of phonon contributions in the frequency domain.

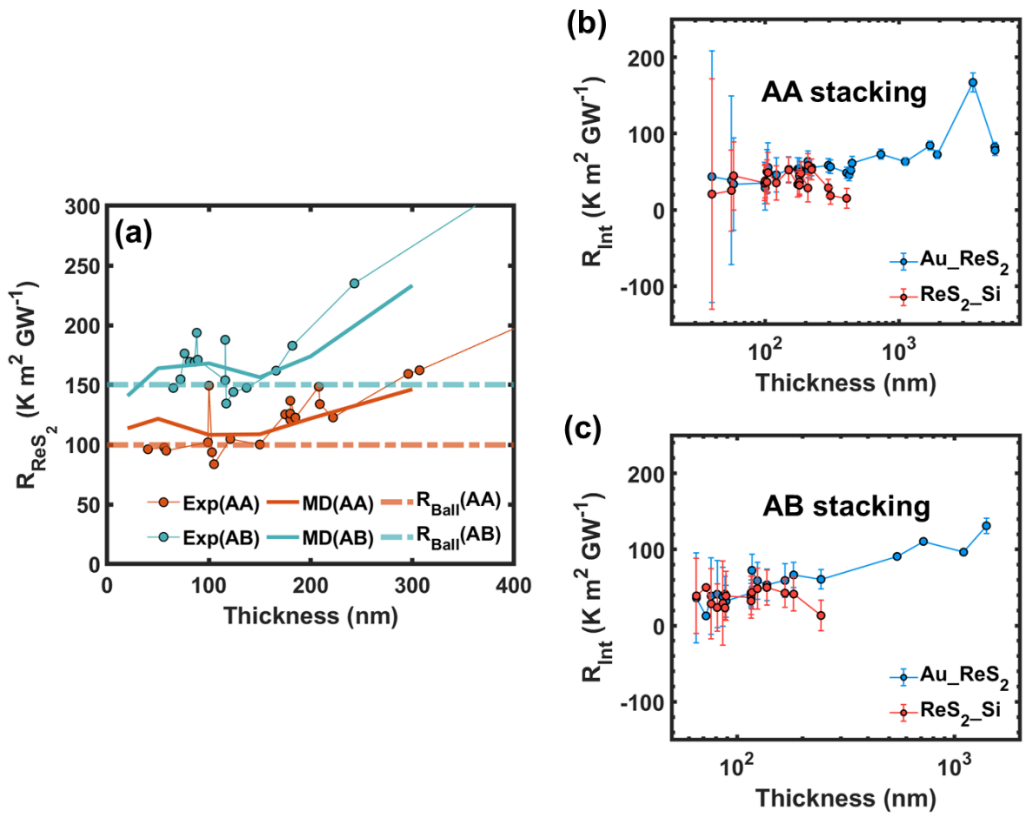

**Figure 5.** (a) Thickness-dependent cross-plane thermal resistance of AA- and AB-stacked $ReS_2$, with corresponding NEMD simulation results. (b) Thickness-dependent contact resistances at the $Au/ReS_2$ and $ReS_2/Si$ interfaces for AA-stacked samples. (c) Thickness-dependent contact resistances at the $Au/ReS_2$ and $ReS_2/Si$ interfaces for AB-stacked samples.

To further elucidate the thickness dependence of cross-plane heat transport, the measured $\kappa_c$ was converted to thermal resistance using $R=t/\kappa_c$, where $t$ is the film thickness. As shown in Fig. 5a, the resistance of ReS$_2$ does not decrease linearly with thickness, but instead exhibits a sublinear, saturating trend as thickness decreases, a hallmark of quasi-ballistic transport [13, 31], in which phonon MFPs are comparable to the film thickness. In the thinnest flakes (<150 nm), both AA- and AB-stacked samples display a clear plateau in resistance, indicating that the heat flow has transitioned into the fully ballistic regime where the conductance becomes thickness-independent. [32, 33] This behavior is faithfully reproduced by our NEMD simulations. From the magnitude of the plateau, we extract ballistic thermal resistances of $R_{ball} \approx 150$ m$^2$ K GW$^{-1}$ for AB stacking and $\approx 100$ m$^2$ K GW$^{-1}$ for AA stacking, values an order of magnitude larger than those reported for MoS$_2$ (~10 m$^2$ K GW$^{-1}$) [13]. Fig. 5b & 5c present the corresponding thickness-dependent contact resistances for both stacking orders. The contact resistances show no systematic dependence on thickness, and the values at the Au/ ReS$_2$ interface (~50 m$^2$ K GW$^{-1}$, or 20 MW m$^{-1}$ K$^{-1}$) agree well with previously reported TDTR measurements and further confirms that the thickness dependent trends do come from intrinsic thermal transport in $R_{ReS2}$. [8]

The direct experimental observation of the ballistic limit in ReS$_2$ is a notable outcome of this study, enabled by two key factors. First, the ps-TTR technique, with its single-pulse excitation and long delay-time window, captures the intrinsic thermal response of ultrathin films without cumulative heating, allowing early-time ballistic transport to be resolved. Second, the exceptionally long and stacking-dependent cross-plane phonon MFPs in ReS$_2$, arising from its weak and registry-sensitive vdW coupling, ensure that the condition $t \lesssim \Lambda_c$ is reached at experimentally accessible thicknesses, permitting the ballistic resistance plateau to emerge directly in the measurements. The resulting finite intercept $R_{ball}$ represents the intrinsic ballistic resistance, reflecting the fundamental phonon transmission across atomic layers and setting the upper bound for heat dissipation in ultrathin ReS$_2$. The substantially larger $R_{ball}$ in ReS$_2$ compared with MoS$_2$ indicates a much lower intrinsic phonon transparency across its vdW gaps, consistent with a narrower spectral window that admits only low-frequency, long-wavelength phonons. The contrast between AA and AB stacking demonstrates that stacking order can serve as a practical tuning parameter for $R_{ball}$, enabling control over interlayer phonon transmission by modifying registry and coupling strength. Such tunability offers a new approach to engineering vdW thermal filters, phononic interfaces, and low-conductance barriers for 2D electronics and optoelectronic devices.

In summary, this work demonstrates that stacking order, an aspect largely neglected in prior studies of vdW materials, is a key determinant of cross-plane phonon thermal transport in ReS$_2$, with implications for thermal transport engineering in ReS$_2$ and other vdW materials. By integrating systematic ps-TTR measurements on micron-scale crystals with single stacking order, and deep-learning-based molecular dynamics and SED analysis, we demonstrate that both AA and AB stacking orders support exceptionally long phonon MFP and exhibit distinct transitions from quasi-ballistic to fully ballistic transport as sample thickness decreases below ~150 nm. Notably, AA-stacking ReS$_2$ exhibits significantly higher $\kappa_c$ than its AB-stacking counterpart, which is attributed to longer acoustic phonon lifetimes enabled by a more "coherent" interlayer registry of AA stacking structure. Furthermore, pressure-dependent analyses reveal that phonon filtering in ReS$_2$ is a fundamentally frequency-selective mechanism, where weak vdW coupling enforces a low-pass filter that is highly tunable through interlayer coupling. These findings reconcile the distinct definitions of "coherence" held by the heat transfer and condensed matter physics communities by demonstrating that ReS$_2$ inherently exhibits both wave- and particle-like transport characteristics. We show that weak interlayer coupling at ambient pressure maintains phase correlation among long-wavelength phonons over hundreds of nanometers, whereas stronger coupling at high pressure increases the MFP primarily by enhancing phonon group velocities. By bridging the gap between the mean-free-path and coherence-length frameworks, serves as a unified platform for the spectral engineering of thermal transport across diverse vdW materials.

## AUTHOR INFORMATION

**Corresponding Authors**

1. Yaguo Wang: Walker Department of Mechanical Engineering, The University of Texas at Austin, Austin, TX, 78712, USA; Email: yaguo.wang@austin.utexas.edu

2. Yan Wang: Department of Mechanical Engineering, University of Nevada, Reno, Reno, NV 89557, USA; Email: yanwang@unr.edu

**Authors**

1. Yongjian Zhou: Walker Department of Mechanical Engineering, The University of Texas at Austin, Austin, TX, 78712, USA.

2. Haoran Cui: Department of Mechanical Engineering, University of Nevada, Reno, Reno, NV 89557, USA.

3. Zefang Ye: Walker Department of Mechanical Engineering, The University of Texas at Austin, Austin, TX, 78712, USA.

4. Jun-Fu Lin: Department of Geological Sciences, Jackson School of Geosciences, The University of Texas at Austin, 2305 Speedway Stop C1160, Austin, Texas 78712, USA.

**Acknowledgements**

Yongjian Zhou, Zefang Ye and Yaguo Wang acknowledge the funding support from the National Science Foundation (NSF-CBET award no. 211660). Haoran Cui, and Yan Wang thank the support from the National Science Foundation (NSF-CBET award no. 2211696)

**Notes**

The authors declare no competing financial interest.